\documentstyle[12pt,epsf]{article}
\newcommand{\noi}{\noindent}
\newcommand{\eq}{\begin{equation}}
\newcommand{\en}{\end{equation}}
\newcommand{\eqa}{\begin{eqnarray}}
\newcommand{\ena}{\end{eqnarray}}

\newcommand{\vx}{{\vec x}}
\newcommand{\vy}{{\vec y}}

\newcommand{\ageq}{\mbox{}_{\textstyle \sim}^{\textstyle > }}
\newcommand{\ra}{\rightarrow}

\newcommand{\be}{\begin{equation}}
\newcommand{\ee}{\end{equation}}
\newcommand{\bea}{\begin{eqnarray}}
\newcommand{\eea}{\end{eqnarray}}

\newcommand{\bt}{\beta}
\begin{document}
\hbox{}
\noindent  \hbox{~}\hfill HU Berlin--EP--96/36
\begin{center}
\vspace*{1.5cm}

\renewcommand{\thefootnote}{\fnsymbol{footnote}}
\setcounter{footnote}{1}

{\LARGE Dynamical Wilson fermions and the problem of
the
chiral limit in compact lattice QED}\footnote{Work supported 
by the Deutsche
Forschungsgemeinschaft under research grant Mu 932/1-4 }

\vspace*{0.5cm}
{\large
A.~Hoferichter$\mbox{}^1$,
V.K.~Mitrjushkin$\mbox{}^1$ 
\footnote{Permanent address:
Joint Institute for Nuclear Research, Dubna, Russia},
M.~M\"uller-Preussker$\mbox{}^1$ and

\vspace*{0.15cm}
H.~St\"uben$\mbox{}^2$}\\
\vspace*{0.2cm}
{\normalsize
{$\mbox{}^1$ \em Institut f\"{u}r Physik, Humboldt-Universit\"{a}t
zu Berlin, Germany}\\
{$\mbox{}^2$ \em Konrad-Zuse-Zentrum f\"ur 
Informationstechnik Berlin, Germany}\\
}    

\vspace{1cm}
{\bf Abstract}
\end{center}
We compare the approach to the chiral transition 
line $~\kappa_c(\bt)~$ in quenched and  full
compact lattice QED with Wilson fermions within 
the confinement phase, especially in the pseudoscalar sector
of the theory.
We show that in the strong coupling limit ($\beta =0$)
the quenched theory is a good approximation to the full one,
in contrast 
to the case of $\beta =0.8$.
At the larger $\beta$-value 
the transition in the full theory is inconsistent with the
zero--mass limit of the pseudoscalar particle, thus
prohibiting the definition of a chiral limit.

\section{Introduction and model description}

Chiral symmetry as a major concept in continuum quantum field theory 
has remained a problematic topic in lattice gauge theories over the years.
It is well-known, that for Wilson fermions  
chiral symmetry is explicitly broken in QCD and QED on the 
lattice \cite{wil,lues}. 
Hopefully, it can be recovered by fine-tuning of parameters
in the continuum limit. Then some line $~\kappa_c(\bt)~$ in the phase
diagram is associated with the chiral limit of the theory.
On the other hand, for non-vanishing lattice spacing only a partial 
restoration of chiral symmetry at $~\kappa = \kappa_c(\bt)~$ is 
possible with Wilson fermions \cite{kawa,kasm}. It is 
still an open question, how this mechanism of partial symmetry 
restoration should eventually be integrated into the general conception
of spontaneously broken chiral symmetry. One cannot exclude that
the breakdown of some other symmetry group governs the 
dynamics of the transitions at $~\kappa_c(\bt)~$ (e.g., \cite{aoki}).
Viewed in this light the vanishing of the pseudoscalar `pion'
mass $~m_{\pi}~$ for $~\kappa \ra \kappa_c(\bt)~$
is a necessary but not sufficient condition for probing
the chiral limit. 
Another point which sharpened the look on the chiral limit
in QCD \cite{sharpe,gupta}
is the discussion of `enhanced logs' due to quenching, demonstrating
the r$\hat{{\rm o}}$le of dynamical fermions in
chiral properties of the theory.

In this letter we are concerned with the behavior of fermionic
observables close to $~\kappa_c(\bt)~$ in the confinement
phase of compact QED. We confront the
full theory with its valence approximation. Though 
the non-perturbative regime of compact QED is itself an
interesting subject the close analogy with non-Abelian 
gauge theories makes it also a valuable test ground 
for QCD with Wilson fermions.

In the rest of the section we will introduce 
the model and give the main notations.
Starting point is the 
partition function of $4d$ compact QED which reads as follows

\eq
Z_{QED} = \int [dU] [d\bar{\psi}d\psi] 
          ~{\rm e}^{-S_{W}(U,~{\bar \psi}, \psi)}~,
\en

\noi where $~S_{W}(U, {\bar \psi}, \psi)~$ denotes the 
standard Wilson lattice action

\eq
 S_{W} = S_{G}(U) + S_{F}(U, {\bar \psi}, \psi) 
                                              \label{wa}
\en

\noi consisting of the plaquette action 

\eq
 S_{G}(U) =
\beta \cdot \sum_{x;\mu >\nu}
        \,  \bigl( 1 - \cos \theta_{x;\mu\nu} \bigr) ~,
                                              \label{wag}
\en

\noi and the fermionic part $S_{F}(U, {\bar \psi}, \psi)~$ 

\eqa
S_{F}
& = & \sum_{f=1}^2\sum_{x,y} 
{\bar \psi}_{x}^{f} {\cal M}_{xy} \psi_{y}^{f}
\nonumber \\ \nonumber \\
{\cal M}_{xy} & \equiv & \hat{1} - \kappa \cdot 
\Bigl[ \delta_{y, x+\hat{\mu}} \cdot ( {\hat 1} - \gamma_{\mu})
\cdot U_{x \mu} +
\delta_{y, x-\hat{\mu}} \cdot ( {\hat 1} + \gamma_{\mu}) \cdot
U_{x-\hat{\mu}, \mu}^{\dagger} \Bigr]~,
                                              \label{waf}
\ena

\noi with $~\beta = 1/g^{2}_{bare}$, and 
$~U_{x \mu} = \exp (i \theta_{x \mu} ),~
\theta_{x \mu} \in (-\pi, \pi] ~$ represent the link fields.
The plaquette 
angles $~\theta_{x;\, \mu \nu}~$ in eq.(\ref{wag}) 
are given by 
$~\theta_{x;\, \mu \nu} =
  \theta_{x;\, \mu} + \theta_{x + \hat{\mu};\, \nu}
- \theta_{x + \hat{\nu};\, \mu} - \theta_{x;\, \nu}$.
In the fermionic part of the action $~{\cal M}_{xy}~$ denotes 
Wilson's fermionic matrix with the hopping 
parameter $~\kappa$ and the flavor--index $~f$.

The phase diagram of this model
has been studied in \cite{qed2}. Within the region of
values 
$~0 \le \kappa \le 0.30~$
the existence of four phases has been established.
The line $~\kappa_c (\beta)~$ separates both the confinement phase 
($~0 \le \beta < \beta_0~$ with $~\beta_{0} \simeq 1.01~$)
and the Coulomb phase ($~\beta > \beta_0~$) from two upper phases
(which we called the $3^{rd}$ phase at weak and the $4^{th}$ phase
at strong coupling).
At $\beta =0.8$ it was observed that the scalar
condensate $\langle {\bar \psi}\psi \rangle$ has a discontinuity 
at $\kappa_c$, therefore we called the transition at
this point a $1^{st}$ order transition.
However, the $\kappa$--dependence of the pseudoscalar observables
was not studied in detail. 

The fermionic observables to be discussed are the 
pion norm 

\eq
\langle \Pi \rangle 
 =  \frac{1}{4V} \cdot \Bigl\langle \mbox{Tr} \Bigl( {\cal M}^{-1}
\gamma_{5} {\cal M}^{-1} \gamma_{5} \Bigr) \Bigr\rangle_{G}~,
                                      \label{pinorm}
\en
\noi being a good indicator for small eigenvalues of the fermionic
matrix and the mass of the pseudoscalar particle 
$~m_{\pi}$, which is extracted from the pseudoscalar 
zero-momentum correlator 

\eqa
\Gamma (\tau )
&=& - \frac{1}{N_s^6} \cdot \sum_{\vec{x},\vec{y}}~
\left\langle {\overline \psi} \gamma_5 \psi (\tau ,\vx ) \cdot
{\overline \psi} \gamma_5 \psi (0, \vy ) \right\rangle 
\nonumber \\
 \nonumber \\
& \equiv & \frac{1}{N_s^6} \cdot \sum_{\vec{x},\vec{y}}~
\left\langle \Biggl\{ \mbox{Sp}
\Bigl( {\cal M}^{-1}_{x y} \gamma_5 {\cal M}^{-1}_{y x} \gamma_5 \Bigr)
- \mbox{Sp} \Bigl( {\cal M}^{-1}_{x x} \gamma_5 \Bigr) \cdot
  \mbox{Sp} \Bigl( {\cal M}^{-1}_{y y} \gamma_5 \Bigr) \Biggr\} 
\right\rangle_{G}~~.
                              \label{pscorr}
\ena

\noi In eqs.(\ref{pinorm},\ref{pscorr}) 
$~\langle ~~ \rangle_{G}~$ indicates averaging over gauge
field configurations, and $~V=N_{\tau} \cdot N_s^3~$ is the number of sites.
~~Sp~~ means the trace with respect to the Dirac indices.
Other observables like $~\langle \rho_{mon} \rangle~$ -- the density 
of DeGrand-Toussaint monopoles \cite{degr} -- and the
photon correlator have also been determined.

Formally  we define the bare fermion mass parameter $~m_q~$ by
\eq
m_q=\frac{1}{2}\Bigl(\frac{1}{\kappa} - \frac{1}{\kappa_c(\bt)}\Bigr) ~.
                                               \label{mass}
\en
\noi For the determination of the $\kappa_c$ see below.

In this work we discuss data from an $~8^3\times 16~$ lattice at two
values of $~\beta~$ within the confinement phase. For the production of 
dynamical gauge field configurations we employed a CRAY-T3D implementation
of the Hybrid Monte Carlo (HMC) method. A detailed presentation of algorithmic 
issues, like the tuning of the HMC parameters
when approaching $~\kappa_c~$ is deferred to \cite{hmms_prog}.

\section{Effects of dynamical Wilson fermions}

First, we shall discuss the behavior of the bulk observable $~\langle
\Pi \rangle~$ when approaching the line $~\kappa_c(\bt)$
at fixed $~\bt~$ within the range $~0 \leq \bt < \bt_0 $
(confinement phase).  As in our previous work for 
representative $~\bt$--values we have chosen 
$~\bt=0.8~$ and the strong coupling limit
$~\bt=0$.  In the latter limit the comparison with
analytical results is possible (e.g., \cite{kawa,klub}).

Provided the pseudoscalar mass vanishes for 
$~m_q \ra 0~$ it will yield the dominant 
contribution to the pion norm
$~\langle \Pi \rangle \sim 1/m_{\pi}^2$.
In case of a PCAC-like relation between $~m_{\pi}~$ and $~m_q~$ 
the pion norm can be expressed in the following form 

\eqa
\langle \Pi \rangle = \frac{C_0}{m_q} + C_1
~,\quad m_q \ra 0~,
                                            \label{Pole_mpi}
\ena

\noi where $~C_0 > 0~$ -- up to a factor -- is 
the subtracted chiral condensate \cite{maian} (see also \cite{hmm95}).
$~C_1 = ~\langle \Pi_m\rangle ~$ is the contribution of the massive modes.

\begin{figure}[htb]
\vskip -2.5truecm
\epsfysize=570pt\epsfbox{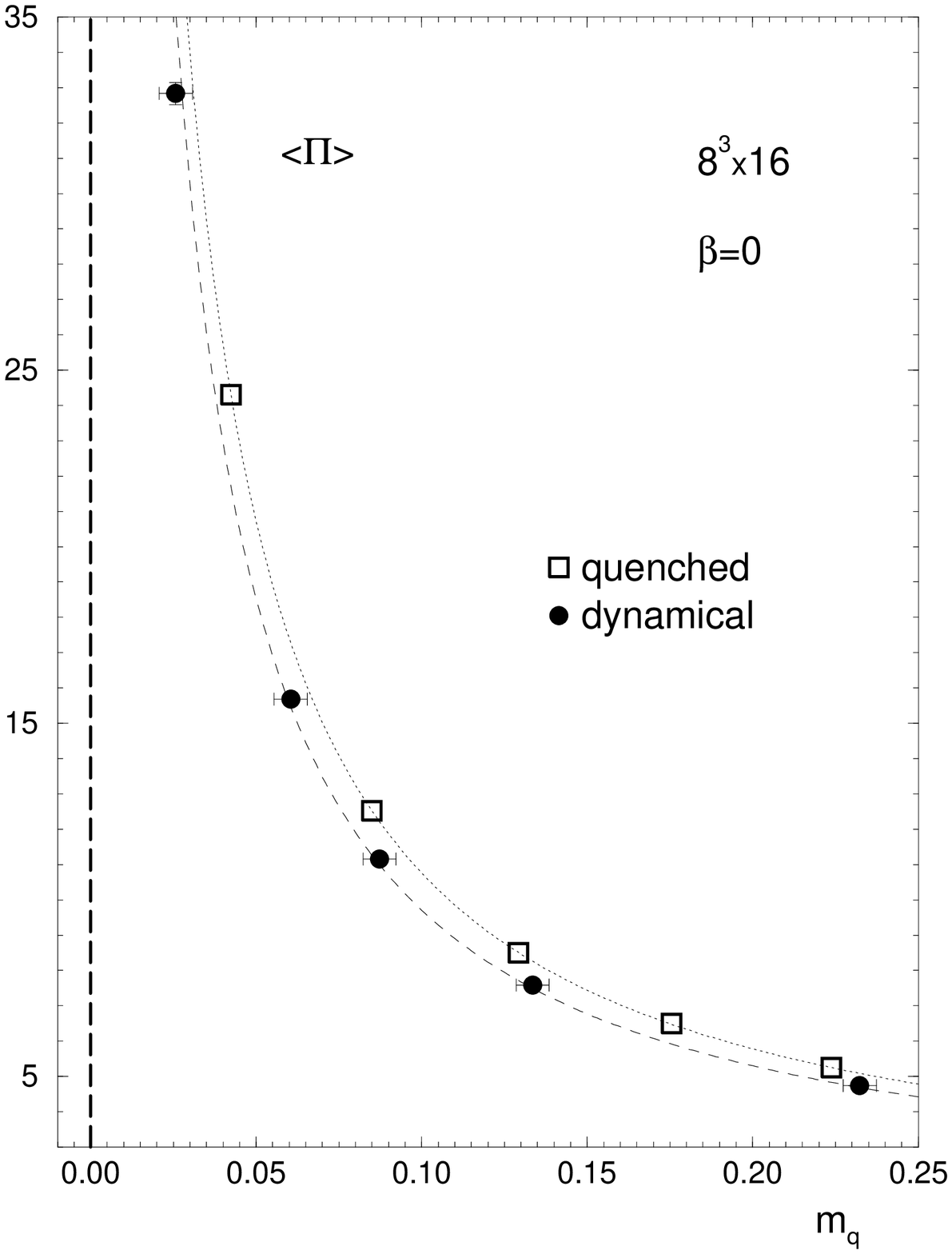}
\caption{
The pion norm $~\langle \Pi \rangle~$ vs. $~m_q~$ at
$~\bt=0$. 
The curved lines correspond to eq.(\protect\ref{Pole_mpi}) 
with $C_0, C_1$ given in Table \protect\ref{tab:one}.
 }
\label{fig:Pi_d_q_b0}
\vskip -0.2truecm
\end{figure}

In Figure \ref{fig:Pi_d_q_b0} data for the quenched 
and dynamical cases at $~\bt=0~$ are presented to view.
The quenched approximation follows the  $~m_q$--dependence
 of the full pion norm $~\langle \Pi \rangle~$ 
very well, even quantitatively.
As $~m_q \ra 0+~$ the singularity of $~\langle \Pi \rangle~$ 
at $~\bt=0~$ is well described in both
cases by eq.(\ref{Pole_mpi}) 
with values of $~C_0,~C_1$ listed in Table \ref{tab:one}.
Note that $C_0~$ and $~C_1~$ differ somewhat for the quenched and
dynamical cases and could not be forced to coincide by shifting 
$~\kappa_c$. The quenched and dynamical theories nevertheless
exhibit the same functional dependence on $~m_q~$ when $~m_q \ra 0$. 

For $\kappa \,\ageq\, \kappa_c~$
the averages of fermionic bulk observables in the quenched
approximation become poorly defined due to large fluctuations caused by
`exceptional' configurations  \cite{hmm95,except}.
In the theory with dynamical fermions we also observe increasing 
fluctuations of, e.g., $~\Pi~$ when $~\kappa~$ is tuned 
towards $~\kappa_c$ from below. 
We were able to  proceed to $~\kappa=0.242~$ ($m_q \sim 0.0253$) 
at $~\bt =0$, before being faced with 
serious problems with the acceptance rate of the HMC method.

\vspace{0.20cm}

\begin{figure}[htb]
\begin{center}
\vskip -2.5truecm
\leavevmode
\hbox{
\epsfysize=570pt\epsfbox{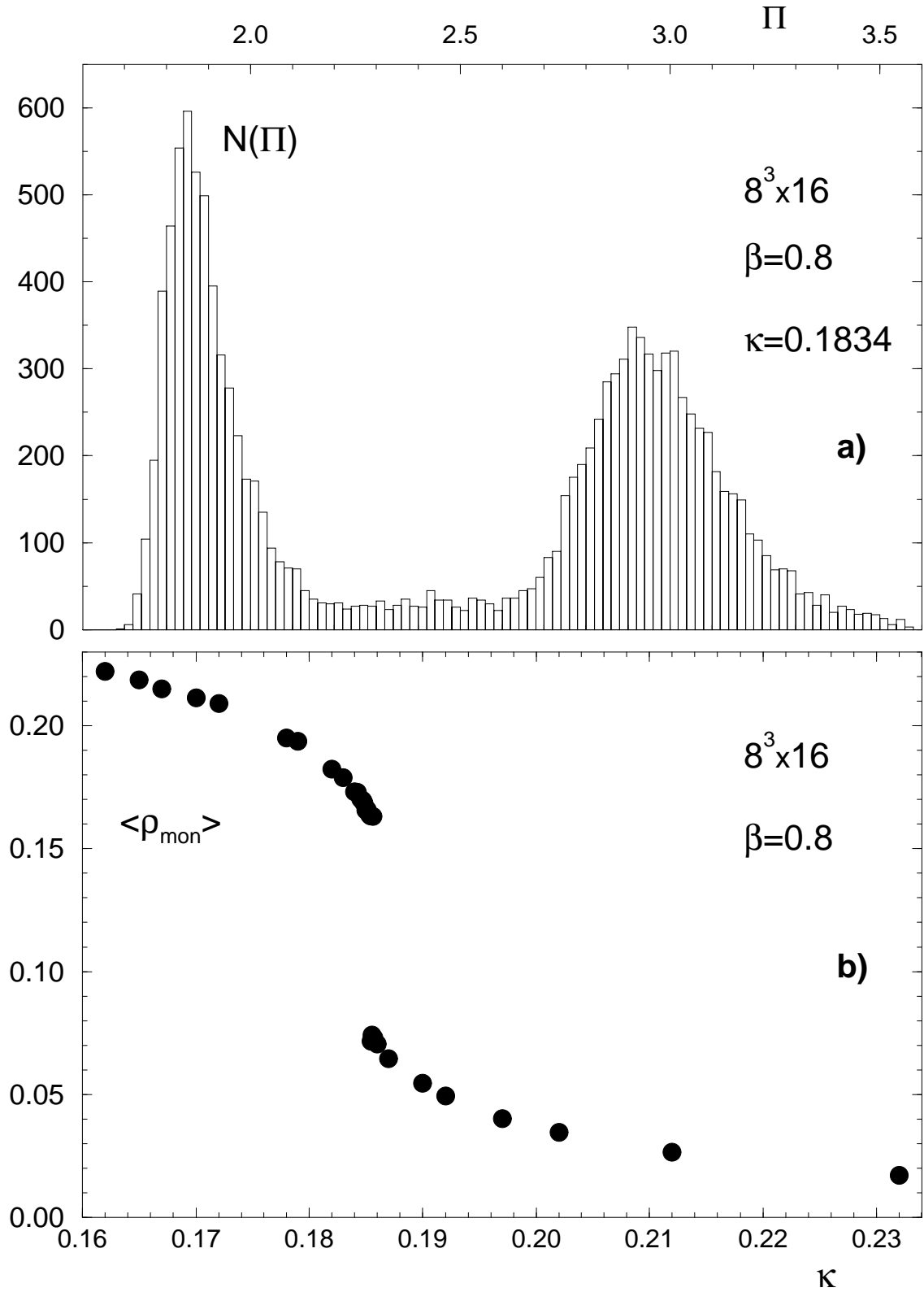}
     }
\end{center}
\caption{
The unnormalized distribution of $~\Pi~$ at $~\bt=0.8~$ in the
vicinity of $~\kappa_c$ ({\bf a}) and $~\langle \rho_{mon} \rangle~$
in dependence of $~\kappa~$ at the same value of $~\bt$ ({\bf b}) both
for the theory with dynamical fermions.
 }
\label{fig:MonDist.d.08}
\vskip -0.2truecm
\end{figure}

The situation changes drastically 
when the gauge coupling $~\bt~$ is increased. 
By examining  time histories of $~\Pi~$ and other observables
at $\beta =0.8$ we observe the formation of metastable states 
in a `critical' region around $~\kappa_c$
in the presence of dynamical fermions.
Figure \ref{fig:MonDist.d.08}a   
illustrates a clearly double peaked distribution of $~\Pi~$ 
close to $~\kappa_c~$ for the case of dynamical fermions.

As an example concerning the behavior of other observables, the
dependence of 
$~\langle \rho_{mon} \rangle~$ on $~\kappa~$ is depicted in Figure
\ref{fig:MonDist.d.08}b.  The evolution of $~\langle \rho_{mon}
\rangle~$ resembles the situation of the confinement--deconfinement
transition at $~\bt_0~$ for a sufficiently small fixed $~\kappa$ and
varying $~\bt~$ \cite{qed2}.

Measurements of the effective photon energy extracted from
plaquette--plaquette correlators for non-zero momentum
confirm that in the case of dynamical fermions the system
undergoes a confinement--deconfinement
transition at $~\kappa_c~$ (see \cite{qed2}).
With increasing $~\kappa~$ the effective
energy of the photon rapidly decreases around $~\kappa_c~$ and becomes
well consistent with the lattice dispersion relation for a zero--mass photon.

We used further `order parameters' to determine 
$\kappa_c$ and to make sure that there are no other transition points
different from that within the investigated $\kappa$--range.
For example, the variance 
$~\sigma^2(\Pi)$ which is a suitable parameter to locate the 
line $~\kappa_c(\bt)~$ within the Coulomb phase \cite{qed2} peaks 
at the same $~\kappa_c$.

\begin{figure}[htb]
\begin{center}
\vskip -2.5truecm
\leavevmode
\hbox{
\epsfysize=570pt\epsfbox{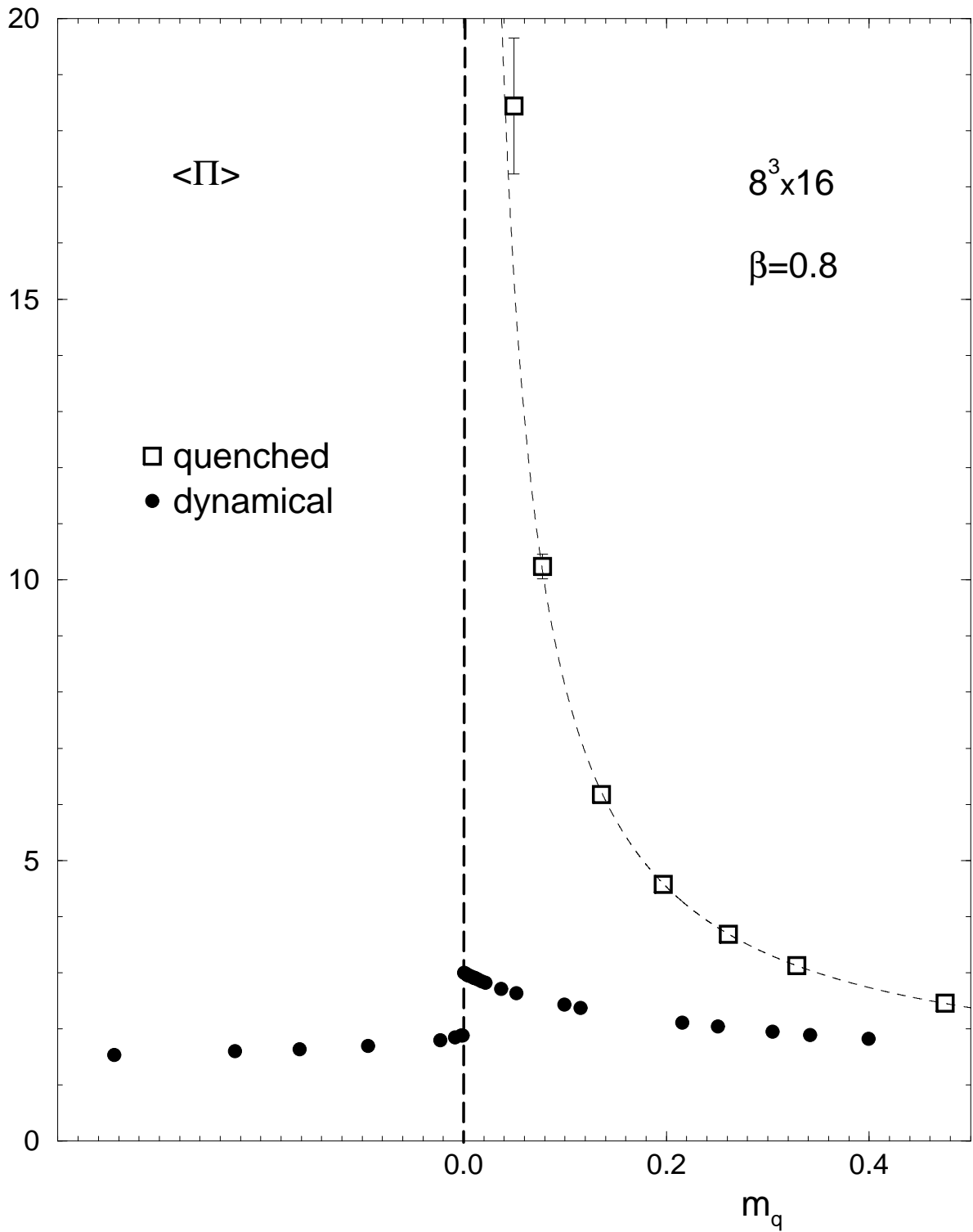}
     }
\end{center}
\caption{
Counterpart to Figure \protect\ref{fig:Pi_d_q_b0} at $~\bt=0.8$.
 }
\label{fig:Pi_d_q_b08}
\vskip -0.2truecm
\end{figure}

In Figure \ref{fig:Pi_d_q_b08} as counterpart of Figure \ref{fig:Pi_d_q_b0}
we confront the dependence of $~\langle \Pi \rangle ~$ on $~m_q~$
in the quenched approximation with the corresponding data when 
dynamical fermions are taken into account. 
The general features of the valence
approximation at $~\bt=0.8~$ are the same as at $~\bt=0$.
$~\langle \Pi \rangle ~$ has a singularity
for $~m_q \ra 0+~$ described very well by eq.(\ref{Pole_mpi})
(dashed line in Figure \ref{fig:Pi_d_q_b08}) with
parameters $~C_0;~C_1~$ given in Table \ref{tab:one}.
However, the effect of the fermionic determinant is now seen
as a qualitative change in the behavior of the pion norm, which
does not behave as $~\sim 1/m_q~$  but rather exhibits a finite 
discontinuity accompanied by a metastable behavior
(cf. Figure \ref{fig:MonDist.d.08}a) around
$~\kappa_c$.
In the quenched theory averages of $~\Pi~$ become statistically not 
well--defined for $~\kappa \,\ageq\, \kappa_c$, as in the case of $~\bt=0$, 
while in the dynamical case at $~\bt=0.8~$ there is no
problem to go beyond $~\kappa_c~$ (i.e., into the 3$^{rd}$ 
phase \cite{qed2}). The dependence of $~\langle \Pi \rangle~$ 
on $~\kappa~$ (respectively $~m_q$) is not symmetric
around $~\kappa_c$.

\begin{figure}[htb]
\vskip -2.5truecm
\epsfysize=570pt\epsfbox{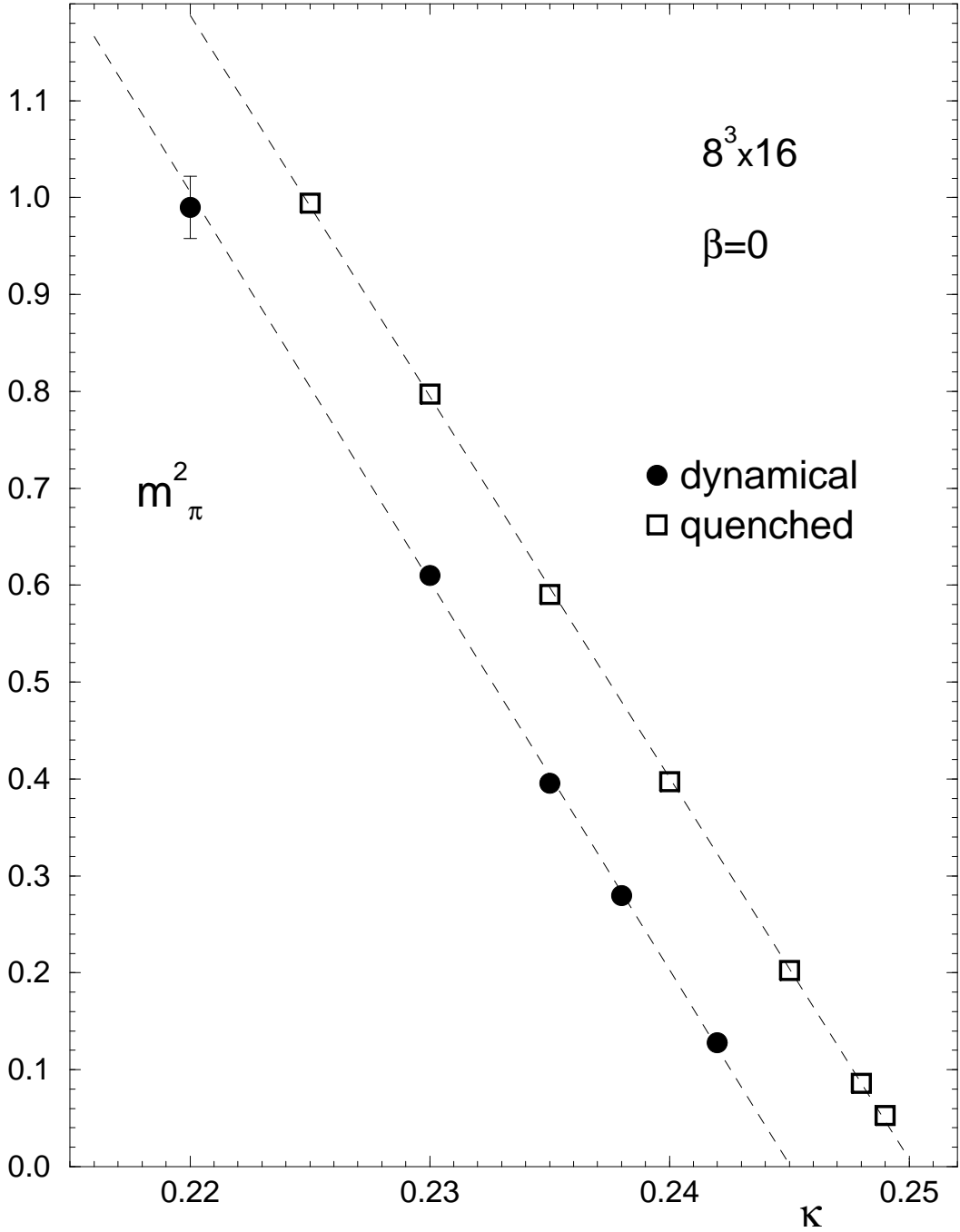}
\caption
{
The $~\kappa$--dependence of $m_{\pi}^2$ for the
quenched and dynamical theories
at $~\bt=0~$ on an $~8^3\times 16~$ lattice.
The broken lines represent linear fits.
 }
\label{fig:Mpi_d_q_b0}
\vskip -0.2truecm
\end{figure}

\vspace{0.25cm}
To substantiate the emerging picture at $~\bt=0~$ and $~\bt=0.8~$
we will discuss the evolution of the pseudoscalar mass 
$~m_{\pi}~$ when $~\kappa \ra \kappa_c$.
We present in Figure \ref{fig:Mpi_d_q_b0} the dependence of 
$~m_{\pi}^2~$ on $~\kappa~$ for the full and 
quenched theories on an $~8^3\times 16~$ lattice 
at $~\bt=0$.
The quenched data for $~\kappa$ very close to $~\kappa_c~$
are obtained by an improved estimator 
of $~m_{\pi}~$ \cite{hmm95} in order to increase the 
signal-to-noise ratio.
Since the existence of a massless pseudoscalar particle is predicted 
by strong coupling arguments \cite{kawa} we extrapolate our
dynamical data for $~m^2_{\pi}~$ linearly to zero in order to determine
$~\kappa_c(\bt=0)~$ as we have done in the quenched case \cite{hmm95}.
In Table \ref{tab:one} we list the values of $~\kappa_c~$ obtained for the 
dynamical and the quenched cases on an $~8^3\times 16~$ 
lattice.
The extrapolated values of $~\kappa_c~$ for the quenched case
are well consistent with the 
prediction at strong coupling \cite{kawa}.
In both, quenched and dynamical cases we observe the following 
dependence of $~m^2_{\pi}~$ on the hopping parameter 
when approaching $~\kappa_c~$ from below 

\eq
m^2_{\pi} \sim \Bigl( 1 - \frac{\kappa}{\kappa_c}\Bigr) ~,\quad 
\kappa \leq \kappa_c~,
                               \label{mpi_k_0}
\en

\noi which in this limit transforms into a 
PCAC--like relation between $~m_{\pi}^2~$ and the bare 
fermion mass $~m_q~$: 
\eq
 m_{\pi}^2 = B \cdot m_q ~,\quad m_q \ra 0+~. 
                                \label{pcac}
\en
The corresponding slopes $~B~$ for the quenched 
and full theories coincide within the errorbars (see Table \ref{tab:one}).
Thus, Figure \ref{fig:Mpi_d_q_b0} suggests a zero-mass pseudoscalar 
particle to exist. 
However, this is a necessary but
not a sufficient prerequisite for the definition of the chiral limit. 

\setlength{\arrayrulewidth}{0.28mm}
\setlength{\doublerulesep}{1.5mm}
\renewcommand{\arraystretch}{1.3}
\begin{table}[htbp]
\centering
\begin{tabular}{|c|c|c|c|c|}\hline
\multicolumn{5}{|c|}{\bf $~\beta=0~$}\\ \hline \hline
{ } & {$\kappa_c$} & {$B$} & {$C_0$} & {$C_1$} \\ \hline
         dynamical & 0.2450(6) &  4.87(5) & 0.895(1) & 0.88(1) \\ \hline
         quenched  & 0.2502(1) &  4.91(4) & 0.996(2) & 0.80(1) \\ \hline\hline
\multicolumn{5}{|c|}{\bf $~\beta=0.8~$}\\ \hline \hline
         dynamical & 0.1832(3) &          &          &         \\ \hline
         quenched  & 0.2171(1) &  3.42(3) & 0.94(2) & 0.72(1) \\ \hline
\end{tabular}
\caption{\label{tab:one}
Compilation of different parameters (see text) for the dynamical 
and quenched theories at $~\bt=0~$ and $~\bt=0.8~$ on 
an $~8^3\times 16~$ lattice.
}
\end{table}

\begin{figure}[htb]
\begin{center}
\vskip -2.5truecm
\leavevmode
\hbox{
\epsfysize=570pt\epsfbox{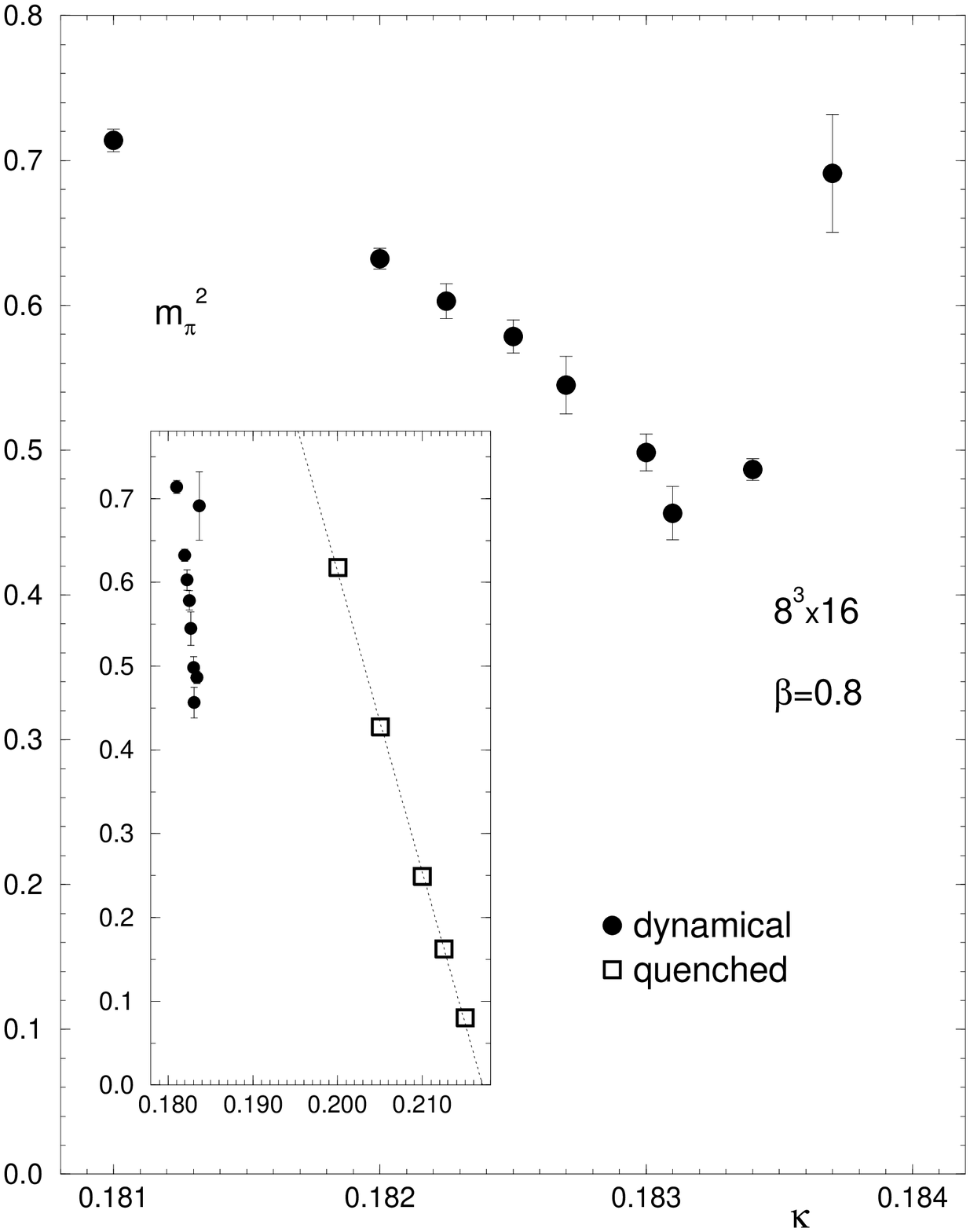}
     }
\end{center}
\caption{
The behavior of $m_{\pi}^2$ around $\kappa_c$ 
at $~\bt=0.8$.
The $~\kappa$-scale of the smaller plot is condensed in order
to permit a direct comparison between the
dynamical and quenched data.
The straight line corresponds to a linear fit of the quenched data.
 }
\label{fig:Mpi_d_b08}
\vskip -0.2truecm
\end{figure}

The corresponding behavior of $~m^2_{\pi}~$ at $~\bt=0.8~$
for quenched and dynamical fermions is
plotted in Figure \ref{fig:Mpi_d_b08}. From the inset of 
Figure \ref{fig:Mpi_d_b08} it can be seen, that as long as the quenched 
theory is considered the situation is fully compatible with 
eq.(\ref{mpi_k_0}). Thus eq.(\ref{pcac}) holds as in 
the case of $~\bt=0$.
Concerning dynamical fermions, again the situation at 
$~\bt=0.8~$ appears to be in sharp contrast to the $~\bt=0$ and to the
quenched cases, as
could be expected from the properties of the pion norm.
By approaching the `critical' value $~\kappa_c(\bt)~$ 
from below with dynamical fermions, the $\kappa$--dependence of $~m^2_{\pi}~$ 
is not linear anymore, i.e. is not compatible with eq.(\ref{mpi_k_0}).
Moreover, 
close to $~\kappa_c(\bt)~$
$m_{\pi}~$ 
has a comparatively large finite minimal value 
which would imply that a zero-mass pseudoscalar particle is not
contained in the spectrum of the theory at this particular coupling.
Increasing $~\kappa~$ beyond $~\kappa_c~$ the pseudoscalar mass starts 
to rise again. Note that in the vicinity 
of $~\kappa_c~$ the dependence of the pseudoscalar mass 
on $~\kappa~$ is different for $~\kappa > \kappa_c~$ 
and $~\kappa < \kappa_c$, in accordance with the discussion 
of the pion norm before.

%
%
%
%
%

%
%
%
%

%
%
%

%
%
%
%
%

%
%

\section{Conclusions and discussion}

We have studied the approach to $~\kappa_c(\beta)~$ for two
$~\bt$--values within the confinement phase of the 
compact lattice QED with Wilson fermions comparing the 
full theory with its quenched approximation. 
We have shown the importance of vacuum polarization effects 
due to dynamical fermions in the context of the chiral limit.

\vspace{0.2cm}
     In the strong coupling limit $~\bt=0$ the 
     main effect of dynamical fermions seems to be a renormalization 
     of the `critical' value $~\kappa_c$,~$~\kappa_c^{dyn} 
     \not = \kappa_c^{quen}$.
     The functional dependence of the studied observables
     on $ ~\kappa~$ (respectively $m_q$) in the limit $~\kappa \ra \kappa_c~$ 
     compared to the quenched approximation does not change.
     Our data suggest, that at $~\bt=0~$ the pseudoscalar particle 
     becomes massless when $~\kappa \ra \kappa_c$.

\vspace{0.2cm}
      At $~\beta=0.8~$ the presence of the dynamical (`sea') fermions
      drastically change the transition.
      There we have found
      a transition which cannot be associated with the zero-mass limit of
      a pseudoscalar particle anymore, in sharp contrast to the quenched case.

\vspace{0.2cm}
      Naively, one would expect that the chiral limit could be
      established everywhere in the confinement phase when approaching the
      `critical' line $\kappa_c(\beta )$. This is not the case.
      Therefore the question arises whether even at $\beta =0$ the
      vanishing pseudoscalar mass can be interpreted as a chiral limit.
	An alternative scenario for the vanishing pseudoscalar mass could be 
      the breakdown of some other symmetry than the chiral one.

\vspace{0.2cm}
	Another possible conclusion from these observations 
      is that the calculations -- analytical 
      and numerical -- in the strong coupling limit 
      $~\beta =0$ can hardly serve for the interpretation of the
      mechanism of the chiral transition at larger $~\beta$'s
      (at least, in this model).

\vspace{0.2cm}

\noi These statements need further confirmation, especially on larger
lattices.  Here work is in progress \cite{hmms_prog}.
It is interesting to what extent the given conclusions can be generalized
to the case of QCD.

\vspace{1.0cm}
\noi {\Large {\bf  Acknowledgement}}

\vspace{0.5cm}
V.K.M. acknowledges support by EEC--contract CHRX-CT92-0051
during his stay at Swansea University where part of this work was done.
The calculations have been performed on CRAY-T3D, CRAY-YMP, CRAY-J90 at
Konrad--Zuse--Zentrum Berlin and on Convex--C3820 at the computer center of 
Humboldt University Berlin.


\vspace{1cm}

\noi {\LARGE {\bf  Figure captions}}

\vspace{0.5cm}
\noi {\bf Figure \ref{fig:Pi_d_q_b0}:}
The pion norm $~\langle \Pi \rangle~$ vs. $~m_q~$ at
$~\bt=0$. 
The curved lines correspond to eq.(\protect\ref{Pole_mpi}) 
with $C_0, C_1$ given in Table \protect\ref{tab:one}.

\vspace{0.25cm}
\noi {\bf Figure \ref{fig:MonDist.d.08}:} 
The unnormalized distribution of $~\Pi~$ at $~\bt=0.8~$ in the
vicinity of $~\kappa_c$ ({\bf a}) and $~\langle \rho_{mon} \rangle~$
in dependence of $~\kappa~$ at the same value of $~\bt$ ({\bf b}) both for 
the theory with dynamical fermions.

\vspace{0.25cm}
\noi {\bf Figure \ref{fig:Pi_d_q_b08}:}
Counterpart to Figure \protect\ref{fig:Pi_d_q_b0} at $~\bt=0.8$.

\vspace{0.25cm}
\noi {\bf Figure \ref{fig:Mpi_d_q_b0}:}
The $~\kappa$--dependence of $m_{\pi}^2$ for the
quenched and dynamical theories
at $~\bt=0~$ on an $~8^3\times 16~$ lattice.
The broken lines represent linear fits.

\vspace{0.25cm}
\noi {\bf Figure \ref{fig:Mpi_d_b08}:}
The behavior of $m_{\pi}^2$ around $\kappa_c$ 
at $~\bt=0.8$.
The $~\kappa$-scale of the smaller plot is condensed in order
to permit a direct comparison between the
dynamical and quenched data.
The straight line corresponds to a linear fit of the quenched data.


\vspace{1cm}

\begin{thebibliography}{99}

\newcommand{\prd}[1]{Phys.~Rev.~{\bf D#1}\ }
\newcommand{\plb}[1]{Phys.~Lett.~{\bf #1B}\ }
\newcommand{\npb}[1]{Nucl.~Phys.~{\bf B#1}\ }
\newcommand{\prl}[1]{Phys.~Rev.~Lett.~{\bf #1}\ }
\newcommand{\pr}[1]{Phys.~Rep.~{\bf #1}\ }
\newcommand{\ap}[1]{Ann.~Phys.~{\bf #1}\ }
\newcommand{\cmp}[1]{Commun.~Math.~Phys.~{\bf #1}}
\newcommand{\rmp}[1]{Rev.~Mod.~Phys.~{\bf #1}}
\newcommand{\ptp}[1]{Prog.~Theor.~Phys.~{\bf #1}}
%
\bibitem{wil}   K. Wilson, \prd{10} (1974) 2445;
                in New phenomena in subnuclear physics, ed. A. Zichichi
                (Plenum, New York, 1977).
\bibitem{lues}  M. L\"uscher, Commun. Math. Phys. {\bf 54} (1977) 283.
\bibitem{kawa}  N. Kawamoto, Nucl. Phys. {\bf B190} (1981) 617.
\bibitem{kasm}  N. Kawamoto and J. Smit, Nucl. Phys. {\bf B192} (1981) 100.
\bibitem{aoki}  S. Aoki, \prd{30} (1984) 2653; ~\prl{57} (1986) 3136;
                ~Phys. Lett. {\bf B190} (1987) 140; UTHEP-318 (hep-lat/9509008).
\bibitem{sharpe} S. Sharpe, Nucl. Phys. {\bf B} (Proc. Suppl.) 
		     {\bf 17} (1990) 1990; Phys. Rev. {\bf D41} (1990) 3146.\\
		     C. Bernard and  M. Golterman, Phys. Rev. {\bf D46} (1992) 853.
\bibitem{gupta}  R. Gupta, Nucl. Phys. {\bf B} (Proc. Suppl.)
                {\bf 42} (1995) 85.
\bibitem{qed2} A.~Hoferichter, V.K.~Mitrjushkin, M.~M\"uller\--Preussker,
               Th.~Neuhaus and H.~St\"uben, 
               Nucl. Phys. {\bf B434} (1995) 358. 
\bibitem{degr}  T.A. DeGrand and D. Toussaint, Phys. Rev. {\bf D22} (1980) 2478.
\bibitem{hmms_prog} A.~Hoferichter, V.K.~Mitrjushkin, M.~M\"uller\--Preussker 
                 and H.~St\"uben, in preparation.
\bibitem{klub} J.-M. Blairon, R. Brout, F. Englert and J. Greensite,  \\
               Nucl. Phys. {\bf B180} (1981) 439. \\
               H. Kluberg-Stern, A. Morel and B. Petersson, Nucl. Phys.
                {\bf B215} (1983) 527.
\bibitem{maian} M.~Bochicchio, L.~Maiani, G.~Martinelli, G.~Rossi and
                M.~Testa, \\ \npb{262} (1985) 331.
\bibitem{hmm95} A.Hoferichter, V.K.Mitrjushkin and M.M\"uller-Preussker,
                Nucl. Phys. {\bf B} (Proc. Suppl.) {\bf 42} (1995) 669.\\
                A.Hoferichter, V.K.Mitrjushkin and M.M\"uller-Preussker,
                hep-lat/9506006 \\ to appear in Z. f. Physik C.
\bibitem{except}  Ph.~De Forcrand, A.~K\"onig, K.-H.~M\"utter,
                K.~Schilling and R.~Sommer,
                in: ~~Proc. Intern. Symp. on Lattice gauge theory
                (Brookhaven, 1986), (Plenum, New York, 1987). \\
                Ph.~De Forcrand, R.~Gupta, S.~G\"usken, K.-H.~M\"utter,
                A.~Patel, K.~Schilling and R.~Sommer,
                \plb{200} (1988) 143.
\end{thebibliography}
\end{document}